\documentclass[11pt]{article}
\usepackage[margin=1in]{geometry}
\usepackage{graphicx} 
\usepackage[utf8]{inputenc}
\usepackage[font=small,labelfont=bf]{caption}
\usepackage{float}
\usepackage{subcaption}
\usepackage[numbers,sort&compress,elide]{natbib}
\usepackage{bm}
\usepackage{hyperref}
\usepackage{siunitx}
\usepackage{authblk}  
\usepackage{listings}
\usepackage{xcolor}
\DeclareSIUnit\bar{bar}
\usepackage{amsmath, amssymb}
\usepackage{appendix}
\usepackage{overpic}

\hypersetup{colorlinks=true,citecolor=blue, linkcolor=blue,urlcolor=blue}
\graphicspath{{figures/}}   

\usepackage[acronym]{glossaries}

\newacronym{ac}{AC}{alternating current}
\newacronym{aom}{AOM}{acousto‑optic modulator}
\newacronym{bbr}{BBR}{blackbody radiation}
\newacronym{cad}{CAD}{computer‑aided design}
\newacronym{cst}{CST}{Computer Simulation Technology} 
\newacronym{dc}{DC}{direct current}
\newacronym{ef}{EF}{electric field}
\newacronym{fem}{FEM}{finite element method}
\newacronym{fft}{FFT}{fast Fourier transform}
\newacronym{fpga}{FPGA}{field‑programmable gate array}
\newacronym{fwhm}{FWHM}{full width at half maximum}
\newacronym{gs}{GS}{ground state}
\newacronym{hd}{HD}{homodyne detection}
\newacronym{hed}{HeD}{heterodyne detection}
\newacronym{hv}{HV}{high voltage}
\newacronym{ir}{IR}{infrared}
\newacronym{lm}{LM358}{low‑power dual operational amplifier}
\newacronym{na}{NA}{numerical aperture}
\newacronym{odmr}{ODMR}{optically detected magnetic resonance}
\newacronym{pd}{PD}{photodiode}
\newacronym{psd}{PSD}{power spectral density}
\newacronym{pt}{PT}{Paul trap} 
\newacronym{qpd}{QPD}{quadrant photodiode}
\newacronym{rf}{RF}{radio frequency}
\newacronym{rp}{RP}{Red Pitaya} 
\newacronym{snr}{SNR}{signal‑to‑noise ratio}
\newacronym{uhv}{UHV}{ultra‑high vacuum}
\newacronym{uv}{UV}{ultraviolet}
\newacronym{bfp}{BFP}{back‑focal plane}
\newacronym{zfs}{ZFS}{zero‑field splitting}
\newacronym{isc}{ISC}{intersystem crossing}
\newacronym{pl}{PL}{photoluminescence}
\newacronym{zpl}{ZPL}{zero‑phonon line}
\newacronym{psb}{PSB}{phonon sideband}
\newacronym{esi}{ESI}{electrospray ionization}
\newacronym{mw}{MW}{microwave}
\newacronym{peek}{PEEK}{polyether ether ketone}
\newacronym{bpd}{BPD}{balanced photodetector}
\newacronym{pll}{PLL}{phase‑locked loop}
\newacronym{spcm}{SPCM}{single‑photon counting module}
\newacronym{qm}{QM}{quantum mechanics}
\newacronym{gr}{GR}{general relativity}
\newacronym{sgi}{SGI}{Stern–Gerlach interferometer}
\newacronym[
  longplural = nanodiamonds,
  shortplural = NDs
]{nd}{ND}{nanodiamond}
\newacronym{com}{CoM}{center‑of‑mass}
\newacronym{nv}{NV}{nitrogen‑vacancy}
\newacronym{csl}{CSL}{continuous spontaneous localization}
\newacronym{bsm}{BSM}{beyond Standard Model} 
\newacronym{pi}{PI}{principal investigator}
\newacronym{pis}{PIs}{principal investigators}

\begin{document}
\title{Trapping and cooling of nanodiamonds in a Paul trap under ultra-high vacuum: Towards matter-wave interferometry with massive objects}
\author[1]{Omer Feldman}
\author[1]{Ben Baruch Shultz}
\author[1]{Maria Muretova}
\author[1]{Or Dobkowski}
\author[1]{Yonathan Japha}
\author[1]{David Grosswasser}
\author[1]{Ron Folman}
\date{}

\affil[1]{Ben-Gurion University of the Negev, Department of Physics and Ilse Katz Institute for Nanoscale Science and Technology, Be'er Sheva 84105, Israel}
\maketitle
\glsdisablehyper 

\begin{abstract}
\Gls{qm} and \gls{gr}, also known as the theory of gravity, are the two pillars of modern physics. A matter-wave interferometer with a massive particle can test numerous fundamental ideas, including the spatial superposition principle---a foundational concept in \gls{qm}---in previously unexplored regimes. It also opens the possibility of probing the interface between \gls{qm} and \gls{gr}, such as testing the quantization of gravity. Consequently, there exists an intensive effort to realize such an interferometer. While several approaches are being explored, we focus on utilizing nanodiamonds with embedded spins as test particles which, in combination with Stern–Gerlach forces, enable the realization of a closed-loop matter-wave interferometer in space-time. There is a growing community of groups pursuing this path \cite{whitepaper_cern_xxx}. We are posting this technical note (as part of a series of seven such notes), to highlight our plans and solutions concerning various challenges in this ambitious endeavor, hoping this will support this growing community.
In this work we detail the trapping of a nanodiamond at \qty{1e-8}{\milli\bar}, which is good enough for the realization of a short-duration Stern-Gerlach interferometer. We describe in detail the cooling we have performed to sub-Kelvin temperatures, and demonstrate that the nanodiamond remains confined within the trap even under high-intensity \qty{1560}{\nano\meter} laser illumination. We would be happy to make available more details upon request.

\end{abstract}

\glsresetall

\section{Introduction}

\Gls{qm} is a pillar of modern physics. It is thus imperative to test it in ever-growing regions of the relevant parameter space. A second pillar is \gls{gr}, and as a unification of the two seems to be eluding continuous theoretical efforts, it is just as imperative to experimentally test the interface of these two pillars by conducting experiments in which foundational concepts of the two theories must work in concert.

The most advanced demonstrations of massive spatial superpositions have been achieved by Markus Arndt's group, reaching systems composed of approximately 2,000 atoms\,\cite{Fein2019MoleculeSuperpositions}. This will surely grow by one or two orders of magnitude in the near future. An important question is whether one can find a new technique that would push the state of the art much further in mass and spatial extent of the superposition. Several paths are being pursued \cite{RomeroIsart2017CoherentInflation, Pino2018OnChipSuperconductingMicrosphere, Weiss2021LargeDelocalizationOptimalControl, Neumeier2024FastQuantumInterference, Kialka2022RoadmapHighMassMWI} and we choose to utilize Stern-Gerlach forces.

The \gls{sgi} has, in the last decade, proven to be an agile tool for atom interferometry \cite{Amit2019T3SGInterferometer,Dobkowski2025QuantumEquivalence, Keil2021SternGerlachAtomChip}. Consequently, we, as well as others, aim to utilize it for interferometry with massive particles, specifically, \glspl{nd} with a single spin embedded in their center \cite{Wan2016FreeRamsey,Scala2013SpinInducedMWI,Margalit2021CompleteSGI}.

Levitating, trapping, and cooling of massive particles, most probably a prerequisite for interferometry with such particles, has been making significant progress in recent years. Specifically, the field of levitodynamics is a fast-growing field \cite{GonzalezBallestero2021Levitodynamics}. Commonly used particles are silica spheres. As the state of the art spans a wide spectrum of techniques, achievements and applications, instead of referencing numerous works, we take, for the benefit of the reader, the unconventional step of simply mapping some of the principal investigators; these include Markus Aspelmeyer, Lukas Novotny, Peter Barker, Kiyotaka Aikawa, Romain Quidant, Francesco Marin, Hendrik Ulbricht and David Moore. Relevant to this work, a rather new sub-field which is now being developed deals with \gls{nd} particles, where the significant difference is that a spin with long coherence times may be embedded in the \gls{nd}. Such a spin, originating from a \gls{nv} center, could enable the coherent splitting and recombination of the \gls{nd} by utilizing Stern-Gerlach forces \cite{Margalit2021CompleteSGI, Wan2016FreeRamsey, Scala2013SpinInducedMWI}. This endeavor includes principal investigators such as Tongcang Li, Gavin Morley, Gabriel H\'{e}tet, Tracy Northup, Brian D’Urso, Andrew Geraci, Jason Twamley and Gurudev Dutt.

We aim to start with an \gls{nd} of $10^7$ atoms and extremely short interferometer durations. Closing a loop in space-time in a very short time is enabled by the strong magnetic gradients induced by the current-carrying wires of the atom chip \cite{Keil2016FifteenYearsAtomChip}. Such an interferometer will already enable to test the existing understanding concerning environmental decoherence [e.g., from \gls{bbr}, see Appendix \ref{appendix:Gravitationlly_incduce_quantum_stateBR}], and internal decoherence \cite{HenkelFolman2024UniversalLimitPhonons}, never tested on such a large object in a spatial superposition. As we slowly evolve to higher masses and longer durations (larger splitting), the \gls{nd} \gls{sgi} will enable the community to probe not only the superposition principle in completely new regimes, but in addition, it will enable to test specific aspects of exotic ideas such as the Continuous spontaneous localization hypothesis \cite{Adler2021CSLLayering,Gasbarri2021TestingFoundationsSpace}. As the masses are increased, the \gls{nd} \gls{sgi} will be able to test hypotheses related to gravity, such as modifications to gravity at short ranges (also known as the fifth force), as one of the \gls{sgi} paths may be brought in a controlled manner extremely close to a massive object \cite{Geraci2010ShortRangeForceMSpheres,GeraciGoldman2015ShortRangeNanosphereMWI,Bobowski2024ShortRangeAnisotropic,Panda2024LatticeGravAttraction}. Once \gls{sgi} technology allows for even larger masses ($10^{11}$ atoms), we could test the Diósi–Penrose collapse hypothesis (see Appendix \ref{appendix:Gravitationlly_incduce_quantum_state}) \cite{Penrose2014GravitizationQM,FuentesPenrose2018QuantumStateReductionBEC,Howl2019BECUnification,Tomaz2024CollapseTimeMolecules,Bassi2013CollapseModelsRMP} and gravity self-interaction \cite{HatifiDurt2023HumptyDumpty,Grossardt2021DephasingSemiclassical,AguiarMatsas2024SchrodingerNewtonSG} (e.g., the Schr\"{o}dinger-Newton equation). Here starts the regime of active masses, whereby not only the gravitation of Earth needs to be taken into account. Furthermore, it is claimed that placing two such \glspl{sgi} in parallel will allow probing the quantum nature of gravity \cite{Bose2017SpinEntanglementQG,MarlettoVedral2017GravInducedEntanglement}. This will be enabled by \gls{nd} \gls{sgi}, as with $10^{11}$ atoms the gravitational interaction could be the strongest \cite{VanDeKamp2020CasimirScreening,Schut2023RelaxationQGIM,Schut2024MicronSizeQGEM}.

Let us emphasize that, although high accelerations may be obtained with multiple spins, we consider only an \gls{nd} with a single spin as numerous spins will result in multiple trajectories and will smear the interferometer signal. We also note that working with an \gls{nd} with less than $10^7$ atoms is probably not feasible because of two reasons. The first is that NVs that are closer to the surface than \qty{20}{\nano\meter} lose coherence, and the second is that at sizes smaller than \qty{50}{\nano\meter}, the relative fabrication errors become large, and a high-precision \gls{nd} source becomes beyond reach.

Here we present the technical details of our work on one of the building blocks of such an \gls{nd} \gls{sgi}, specifically, the levitation of a \gls{nd} (fluorescent nanodiamond - \gls{nv} centers at $\sim3$ ppm, \qty{90}{\nano\meter} particle size, Sigma-Aldrich) in a Paul trap at \qty{1e-8}{\milli\bar}. We demonstrate the detailed characterization of the levitated particle, as well as cooling, and show that high-intensity laser illumination can be applied to the \gls{nd} while maintaining stable trapping. This technical note is part of a series of seven technical notes put on the archive towards the end of August 2025, including a wide range of required building blocks for the \gls{nd} \gls{sgi} \cite{Skakunenko_Needle_trap, Givon_ND_fabrication, Muretova_ND_theory, Levi_Quantum_control_NV, Liran_ND_neutralization, benjaminov2025uhvloading}.

\section{Experimental Setup}

\begin{figure}[ht]
\centering
\includegraphics[width=\textwidth]{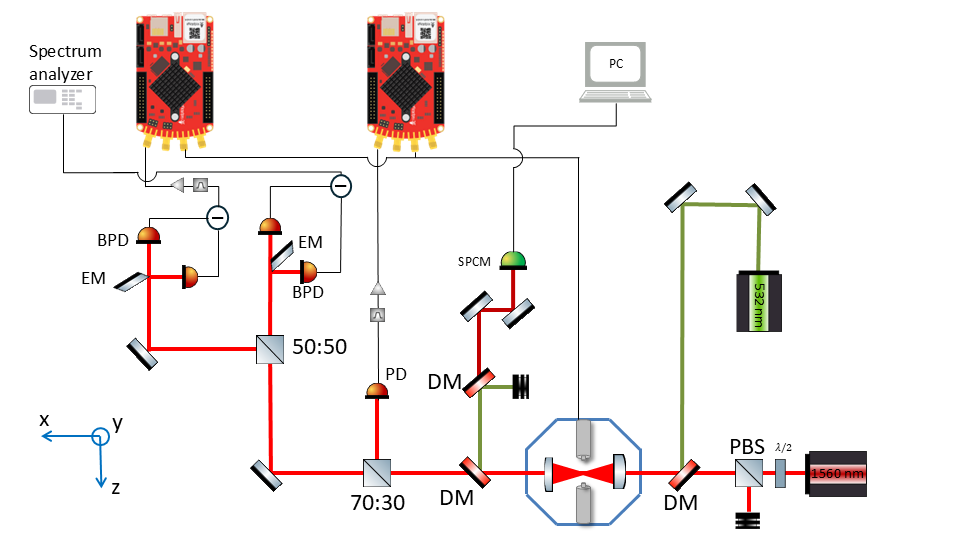}
\caption[Experimental system overview.]{Schematic of the experimental setup. A \qty{1560}{\nano\meter} laser is intensity-controlled using a waveplate and PBS, then co-aligned with a \qty{532}{\nano\meter} laser before passing through the vacuum chamber to illuminate the particle. The outgoing beam is directed into a three-detector split detection array. The \gls{ir} beam provides three-axis position detection, while the \qty{532}{\nano\meter} beam enables \gls{nv} center measurements. EM - edge mirror, BPD - balanced photodetector, PD - photodiode, DM - dichroic mirror, SPCM - single photon counting module, PBS - polarizing beam splitter.}
\label{fig:system_overview}
\end{figure}

We present an experimental system for trapping and analyzing single nanoparticles under \gls{uhv} conditions. The apparatus consists of four main components: an end-cap Paul trap, an \gls{esi} source, a \gls{uhv} chamber with integrated optics, and a homodyne detection system. Fig.\,\ref{fig:system_overview} shows the overall system layout.

\subsection{Paul Trap Design}
The end-cap Paul trap consists of two inner steel rods (\qty{0.5}{\milli\meter} radius) separated by \qty{1}{\milli\meter} and two coaxial cylinders separated by \qty{1.2}{\milli\meter}. The inner electrodes are electrically connected and driven by one \gls{hv} amplifier, while the outer electrodes are similarly connected and driven by a second amplifier. The inner and outer electrodes are driven with opposite \gls{rf} phases. All electrodes are isolated with \gls{peek} sleeves and mounted on a three-axis micrometric stage (TSDS-255S by OptoSigma) driven by piezo actuators (VPDM-6.5 OptoSigma). Small notches cut in the steel jackets break the radial degeneracy.

\begin{figure}[H]
\centering
\begin{subfigure}[t]{0.48\textwidth}
  \centering
  \includegraphics[width=0.95\linewidth]{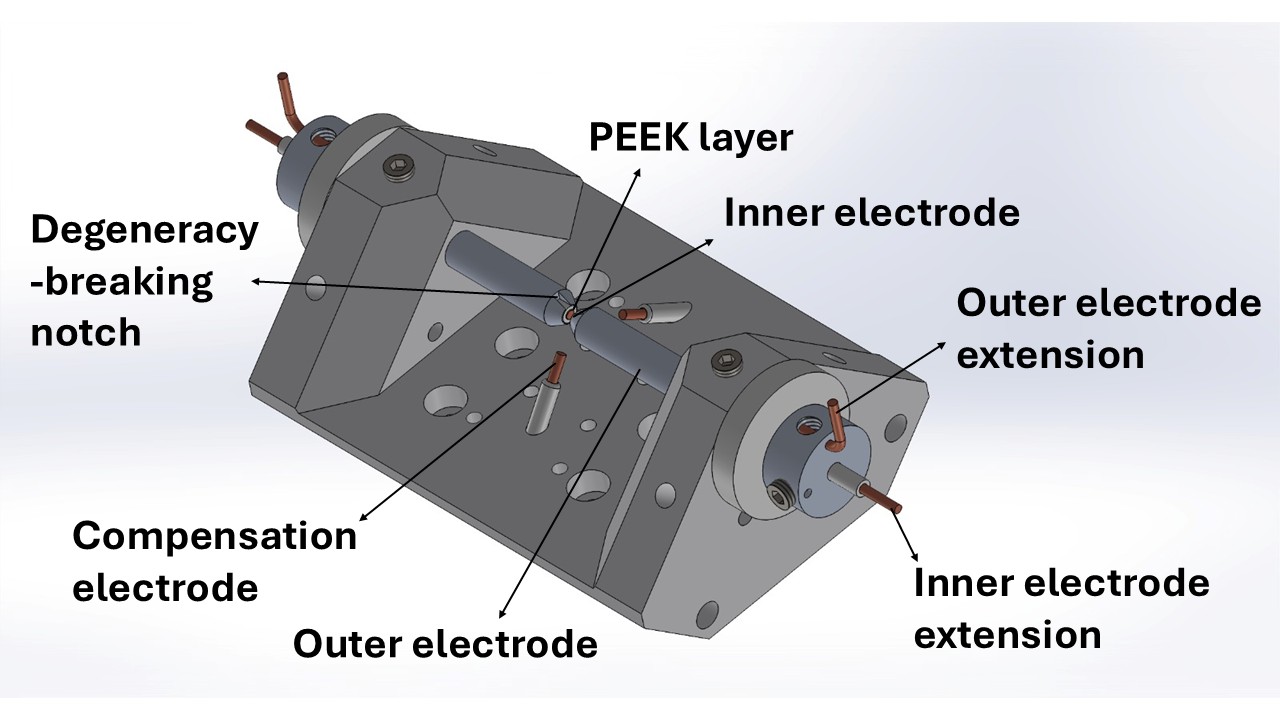}
  \caption{}
  \label{fig:Trap in Solid with labels}
\end{subfigure}
\hfill
\begin{subfigure}[t]{0.48\textwidth}
  \centering
  \includegraphics[width=0.95\linewidth]{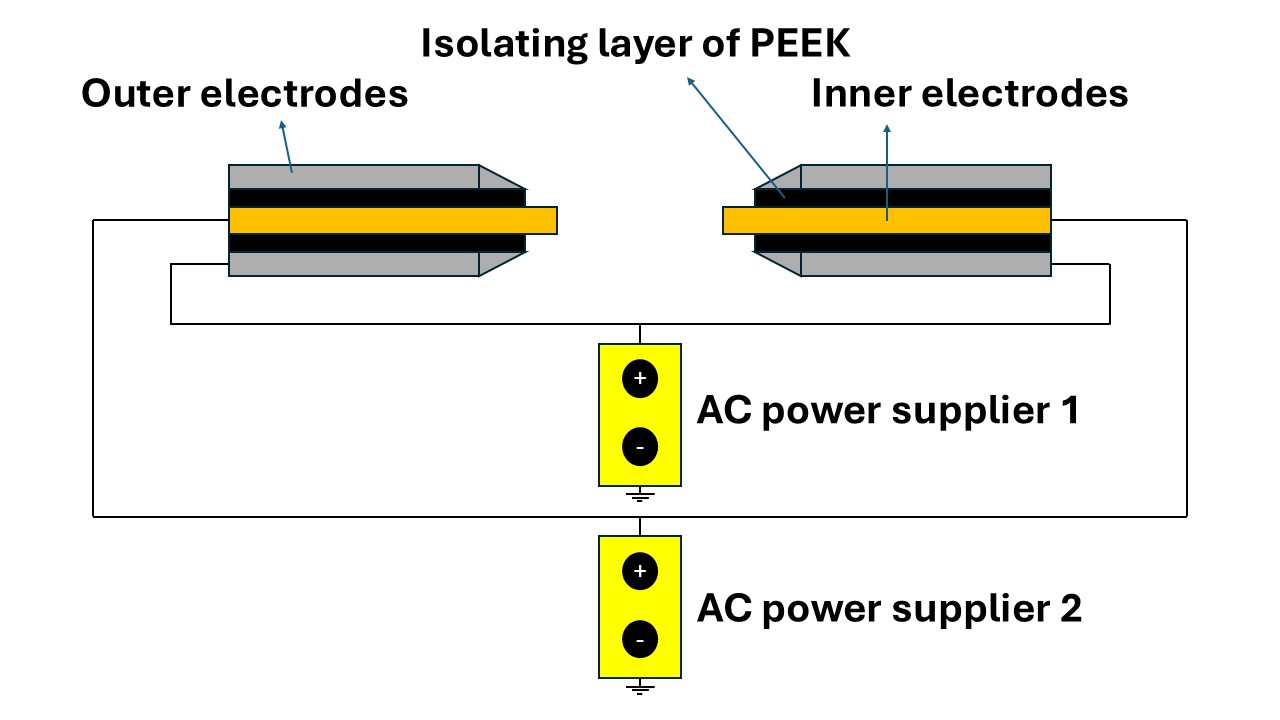}
  \caption{}
  \label{fig:Electrical trapping scheme}
\end{subfigure}

\vspace{0.5em}

\begin{subfigure}[t]{0.65\textwidth}
  \centering
  \includegraphics[width=0.95\linewidth]{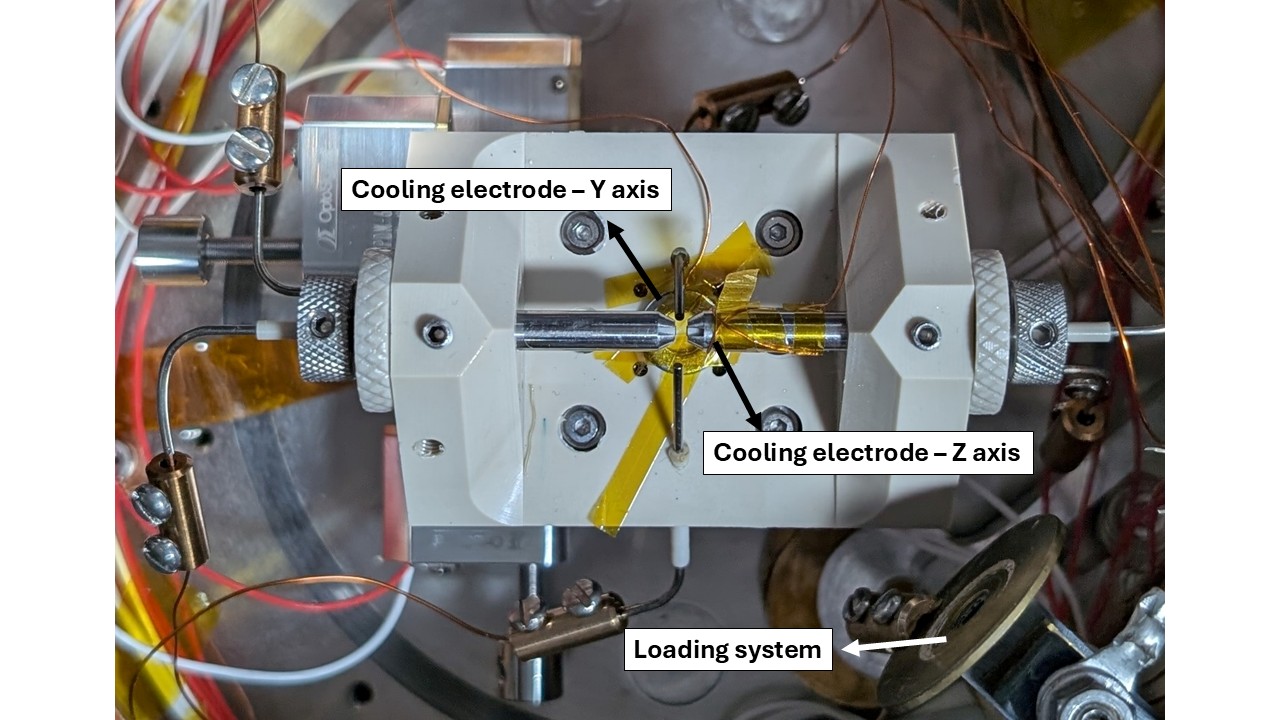}
  \caption{}
  \label{fig:Trap picture with labels}
\end{subfigure}

\caption{
(a) A 3D model of the end-cap trap and its holder made in SolidWorks. The compensation electrodes enable the creation of electric fields that counter stray electric fields in the trap region that may displace the particle from the trap center and create undesired excess micromotion \cite{Berkeland1998MinimizationMicromotion}. (b) A simplified schematic of the electrical connections for the end-cap trap. The AC voltages from the two power supplies are identical in amplitude but shifted by $\pi$ in phase. (c) A photograph of the trap mounted in the vacuum chamber on the 3D piezo stage, viewed from above. The y- and z-axis compensation electrodes are visible (the z-axis electrode consists of a Kapton-wrapped wire looped around the outer electrode), while the x-axis electrode is not shown as it is mounted on the lens assembly (removed before photographing). The yellow material securing the electrodes is Kapton tape. The particle loading system is visible in the bottom right corner.
}
\label{fig:Scheme and 3d model of trapping electrodes}
\end{figure}


The trap operates at \gls{rf} frequencies up to \qty{250}{\kilo\hertz} with amplitudes of \qtyrange{500}{700}{\volt_{pp}} using two \gls{hv} amplifiers (A-304, A.A.~Lab-Systems). For higher amplitudes (\qtyrange{1000}{1600}{\volt_{pp}}), particles are first loaded at reduced voltage, the chamber is evacuated, and then the voltage is increased to avoid voltage breakdown. Direct current (DC) compensation is provided by a programmable power supply (Rigol DP932E).
Three feedback electrodes are connected to control the $x$, $y$, and $z$ axes of \gls{com} motion of the \gls{nd}. 

\subsection{Optical System}

\begin{figure}[H]
    \centering
    \includegraphics[width=0.5\linewidth]{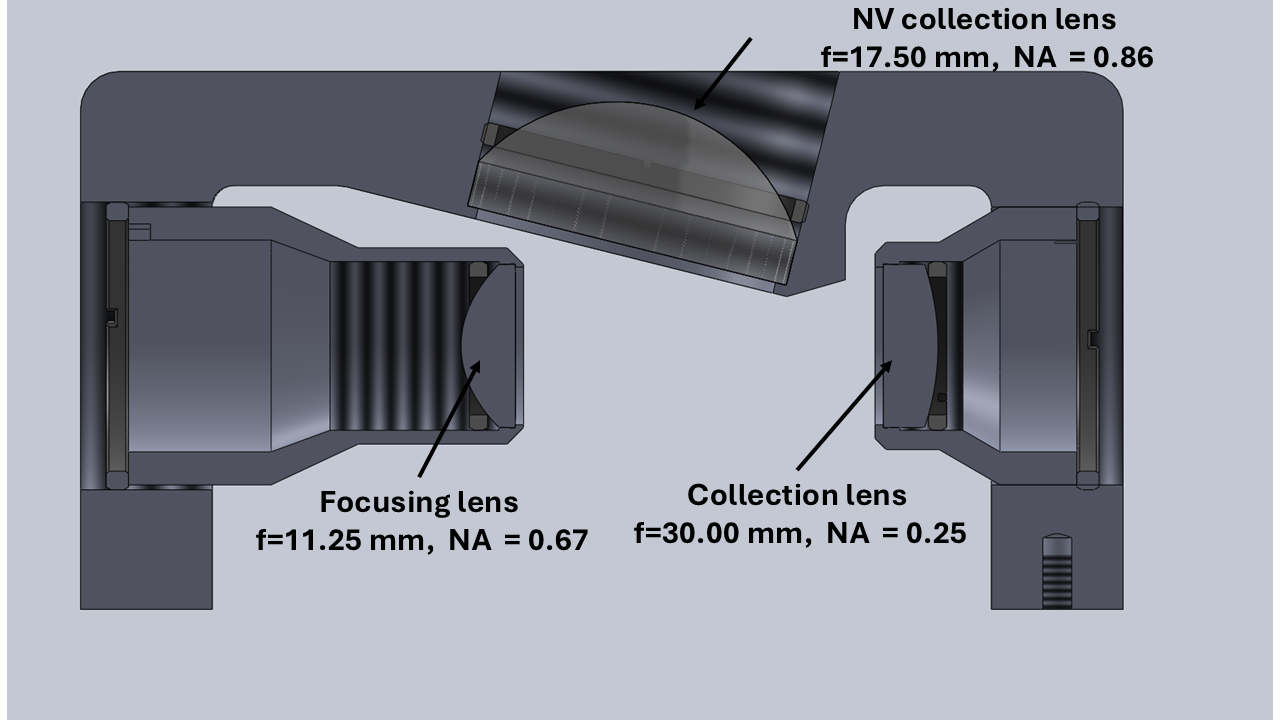}
    \caption{3D model of the lens holder made in SolidWorks. The holder is designed with space to insert a chip for SG forces in future experiments.}
    \label{fig:Lens holder with labels}
\end{figure}

The optical setup employs two co-aligned lasers: a \qty{1560}{\nano\meter} \gls{ir} laser for position detection and a \qty{532}{\nano\meter} laser for \gls{nv} center spectroscopy and imaging. Both beams are focused to approximately \qty{10}{\micro\meter} diameter at the trap center.

The beams are focused inside the chamber using an aspheric lens ($\text{f}=\qty{11.25}{\milli\meter}$, $\text{\gls{na}}=0.67$). The transmitted beam, together with the light scattered from the particle, is collected for forward scattering using an aspheric collection lens ($\text{f}=\qty{30}{\milli\meter}$, $\text{\gls{na}}=0.25$). The \gls{nv} and green optics collection is designed with flexibility to use either a CMOS objective apparatus with a long-distance objective (\qty{150}{\milli\meter} focal length) for imaging applications, or an aspheric collection lens for optimized light collection efficiency. In the current CMOS imaging configuration, the forward detection path simultaneously serves for both position detection and \gls{nv} light collection, as illustrated in Fig.~\ref{fig:system_overview}. 

\begin{figure}[H]
  \centering
  \begin{subfigure}[b]{0.48\textwidth}
    \centering
    \begin{overpic}[width=\linewidth]{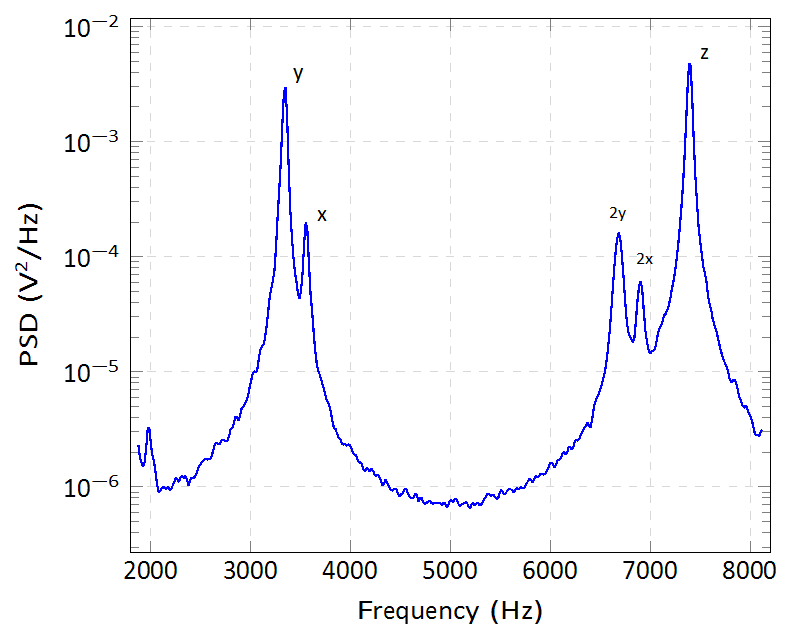}
      \put(25,77){\makebox(0,0)[tr]{\textbf{(a)}}}
    \end{overpic}
    \phantomsubcaption\label{fig:Typical_spectrum}
  \end{subfigure}
  \hfill
  \begin{subfigure}[b]{0.48\textwidth}
    \centering
    \begin{overpic}[width=\textwidth, height=0.8\linewidth]{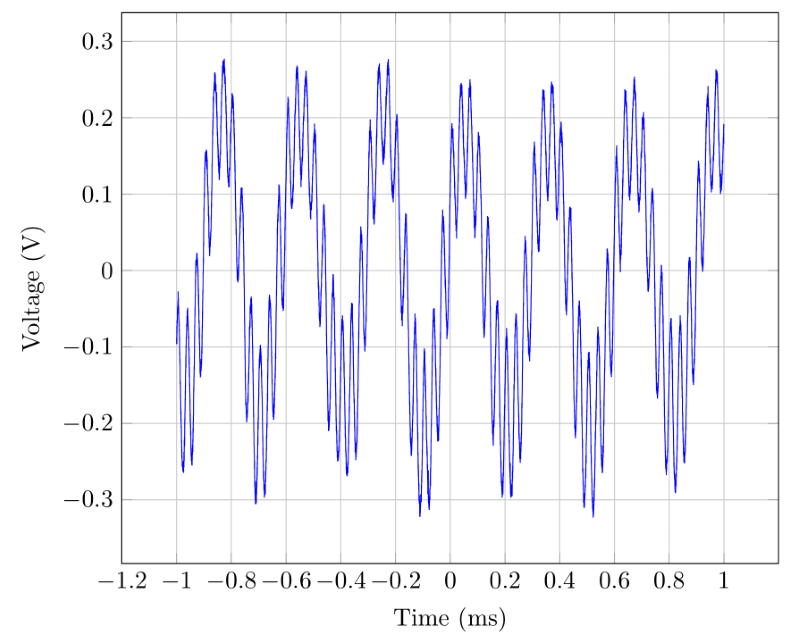}
      \put(24,77){\makebox(0,0)[tr]{\textbf{(b)}}}
    \end{overpic}
    \phantomsubcaption\label{fig:waveform}
  \end{subfigure}
    \caption{Detection: (a) Typical \gls{bpd} \gls{psd} showing the motional resonances of a trapped \gls{nd}: two radial modes ($x,y$) and the axial mode ($z$). Two additional peaks to the left of the axial resonance correspond to the second harmonics of the radial modes. (b) Raw time-domain trace of the detector signal from the oscilloscope, showing the secular motion and micromotion.}
    \label{fig:modes}
\end{figure}

Position detection uses homodyne detection with the forward-scattered \gls{ir} light. After the collection lens, the beam is split into 3 parts for axial and radial channels with beamsplitters, and each channel is further split with D-shaped mirrors and measured using balanced photodetectors (Femto HBPR-100M-60K-IN-FST). The system employs two \glspl{bpd}: one primarily used for detection and feedback (BPD1), and another for out-of-loop measurements (BPD2). The signal from BPD1 is sent to a low-noise voltage preamplifier (SR560 by Stanford Research Systems). Motion spectra are processed in real-time by a \gls{rp} \gls{fpga} board for velocity-dependent feedback cooling. An out-of-loop signal is measured using a spectrum analyzer (SR770 by Stanford Research Systems) for calibration and temperature measurement. A typical spectrum of a single axis of motion can be seen in Fig.~\ref{fig:Typical_spectrum} with a corresponding waveform of motion in Fig.~\ref{fig:waveform}.

Signal calibration relates the measured voltage variance $\langle V^{2}\rangle$ to particle displacement $\langle u^{2}\rangle$ through the equipartition theorem: \begin{equation} \label{eq:calibration_signal} S^{2} = \frac{k_{\mathrm{B}}T}{m\omega^{2}\langle V^{2}\rangle} \end{equation} where $S$ is the conversion factor, $k_{\mathrm{B}}$ is Boltzmann's constant, $m$ is the particle mass, and $\omega$ is the secular frequency.

\subsection{Particle Loading and Vacuum System}
\begin{figure}[ht] 
\centering 
\includegraphics[width=0.7\linewidth]{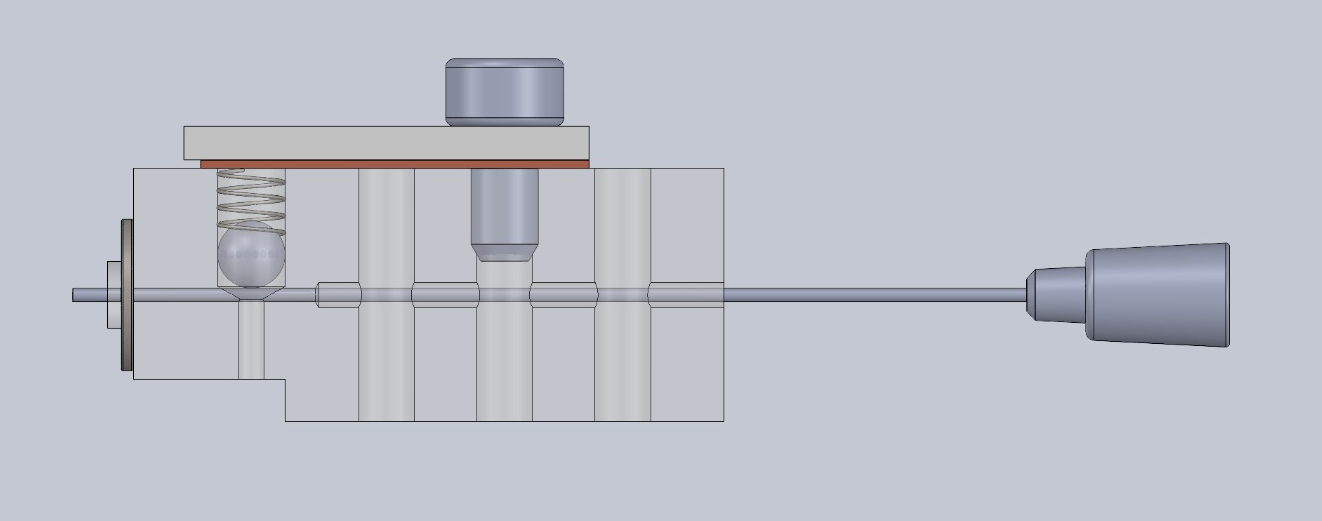} 
\caption[Electrospray internal structure.]{CAD illustration of our homemade \gls{esi}. The needle is inserted into the assembly, creating electrical contact pressed by the steel ball that is pushed by the spring. Applying \gls{hv} to the screw on top provides the necessary potential to generate a spray of charged \glspl{nd} into the trap. A more efficient dry source is being developed in parallel, see our work in \cite{benjaminov2025uhvloading}.}
\label{fig:esi_asem} 
\end{figure}

Nanoparticles are loaded using a custom-made \gls{esi} consisting of a permanent component illustrated in Fig.\,\ref{fig:esi_asem}. The \gls{esi} is connected inside the vacuum chamber to a \gls{hv} power supply (Trek 640-D). A suspension of \qty{90}{\nano\meter} diameter \glspl{nd} in ethanol is sonicated, filtered, and electrosprayed using \qtyrange{1.9}{2.4}{\kilo\volt} applied to a metallic needle. Initially, several particles may be trapped simultaneously. Single particles are then isolated by adjusting the \gls{rf} frequency to eject particles with unfavorable charge-to-mass ratios. The needle is then removed along with the excess solvent. Typical trapping parameters for single particles are \qtyrange{560}{800}{\volt_{pp}} at \qtyrange{25}{100}{\kilo\hertz}.

The vacuum system achieves a vacuum as low as \qty{1e-7}{\milli\bar} through two-stage pumping: a diaphragm backing pump, and a turbomolecular pump (Leybold TurboLab 90, \qty{90}{\liter\per\second}). An ion pump (SAES NEXTorr Z200, \qty{290}{\liter\per\second}) is activated once the pressure is reduced to \qty{1e-6}{\milli\bar} for reducing the pressure down to \qty{1e-8}{\milli\bar}. The vacuum is monitored by a Pirani gauge down to \qty{5e-4}{\milli\bar}. At a higher vacuum, an ion gauge (Kurt J. Lesker Co.\ model KJLC392402YE) is activated, followed by measuring the vacuum using the ion pump current (Nextorr NIOPS-03) when it is activated.


\section{Particle Characterization: Mass and Charge determination}
\label{sec:characterization}

We determined the key physical properties of the trapped \gls{nd}.

A particle can be trapped in a Paul trap only under certain conditions of the Mathieu parameters $q_z$ and $a_z$, which are defined by the Mathieu equations that describe particle motion in a Paul trap. Under no DC fields ($a_z=0$), the condition for stable trapping is:

\begin{equation}
    q_z=\frac{4 Q \eta V_0}{m d^2 \Omega^2} \lesssim 0.908
\end{equation}

where $Q$ and $m$ are the particle's charge and mass, $V_0$ and $\Omega$ are the trap's drive voltage and frequency, $\eta$ and $d$ are geometric factors of the trap.

The procedure for measuring the charge-to-mass ($Q/m$) ratio involves measuring how the axial secular frequency ($\omega_z$) depends on the trap drive voltage. The secular frequency of a particle in a Paul trap is defined by a parameter $\beta_i$ as $\omega_i=\beta_i\Omega/2$, where $\omega_i$ is the secular frequency along the $x$, $y$, and $z$ axes, and $\Omega$ is the drive frequency of the Paul trap \cite{major_charged_2005}. The parameter $\beta$ can be defined by the approximation:

\begin{equation}
    \beta_i^2 \approx a_i +\frac{q_i^2}{2},
    \label{eq:beta_adiabatic}
\end{equation}

\begin{figure}[H]
    \centering
    \includegraphics[width=0.8\textwidth]{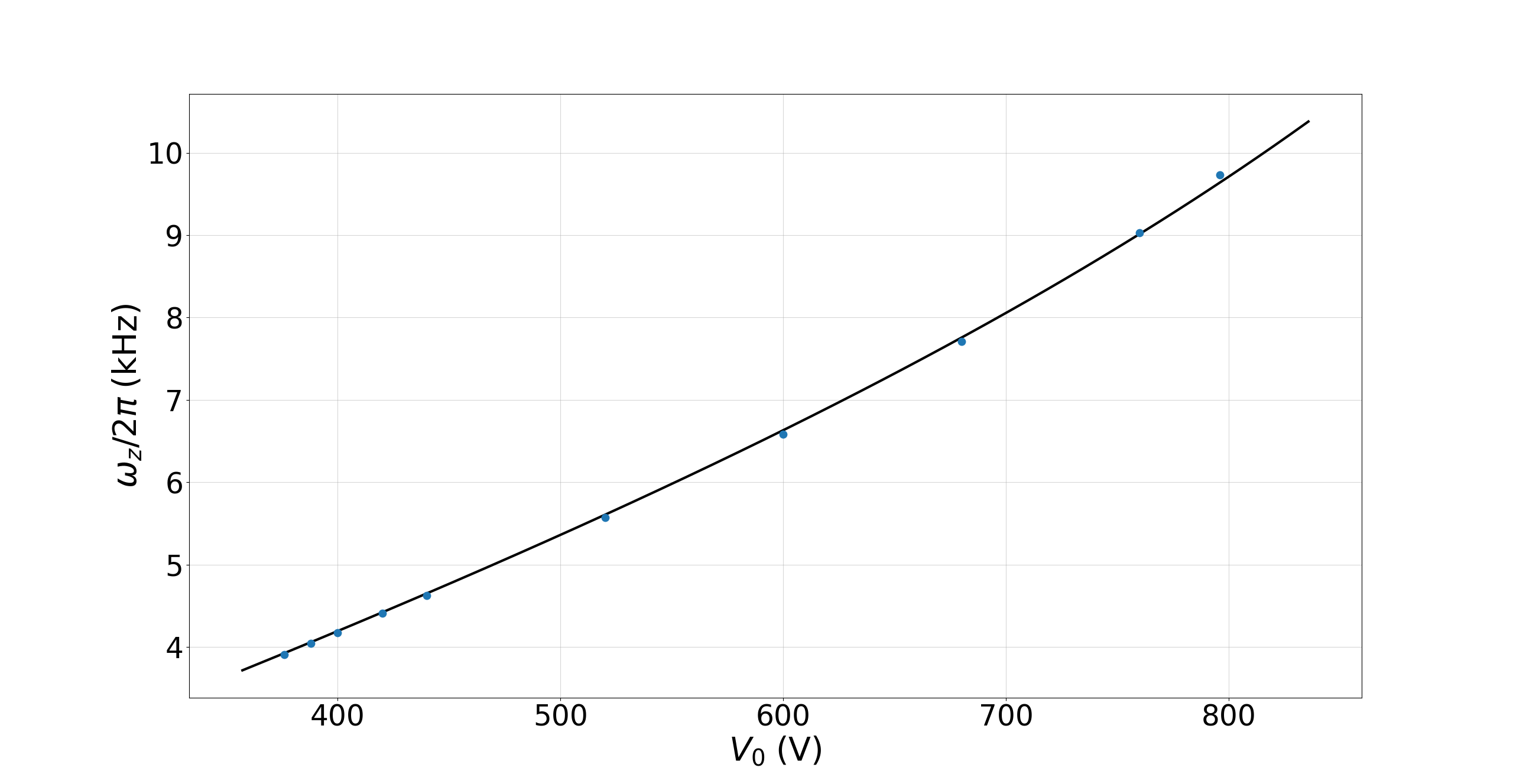}
    \caption[Measurements of ($Q/m$)]{Measurement of the axial secular frequency $\omega_z$ as a function of drive voltage $V_0$ for a trapped \gls{nd} particle. The secular frequency increases with drive voltage due to the voltage dependence of the effective trapping potential. The charge-to-mass ratio ($Q/m$) is extracted by fitting the data to the theoretical relationship $\omega_z = (\beta_z \Omega)/2$, where $\beta_z$ is described by Eq.\,\eqref{eq:beta_adiabatic}. The black solid line shows a fit to $\beta(q_z)$ on the experimental data, yielding $Q/m = $\qty{75}{\coulomb\per\kilogram}.}
    \label{fig:Q_over_m_both}
\end{figure}

By scanning the drive voltage of the trap and measuring $\omega_z$, a fit can be performed to determine the $Q/m$ parameter of the particle. This procedure yields values in the range of \qtyrange{10}{75}{\coulomb\per\kilogram} (Fig.~\ref{fig:Q_over_m_both}).

\begin{figure}[H]
  \centering

  \begin{subfigure}[b]{0.49\textwidth}
    \centering
    \begin{overpic}[width=1.1\linewidth, height=0.7\linewidth]{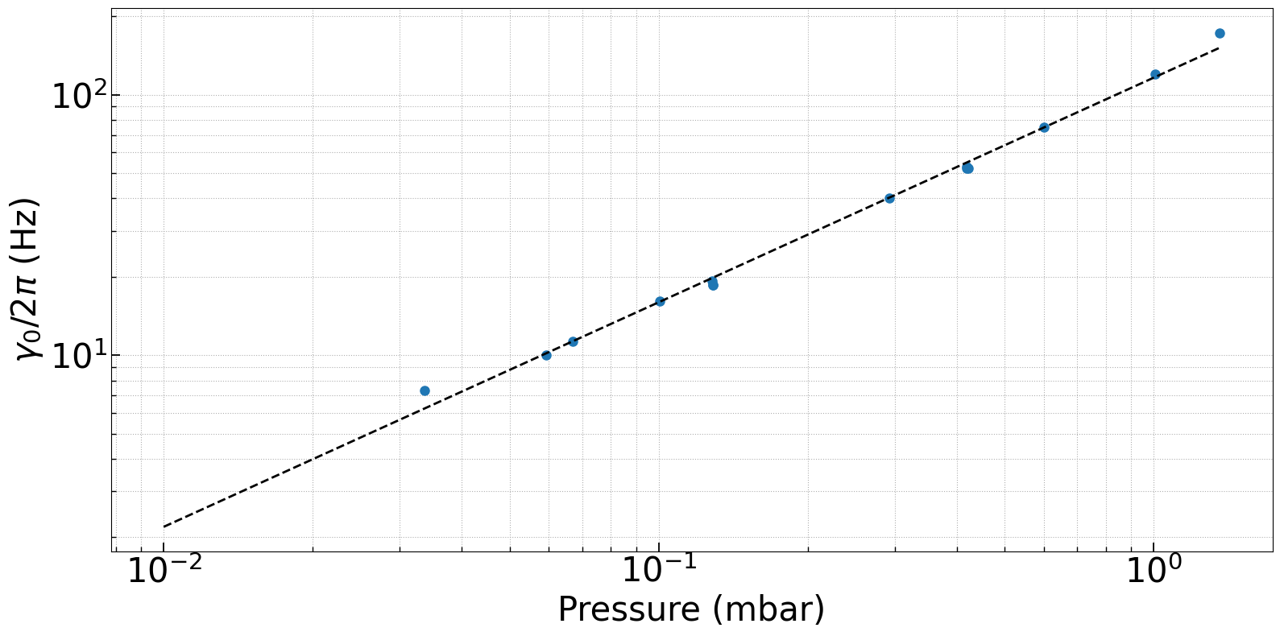}
      \put(16,61){\makebox(0,0)[tr]{\textbf{(a)}}}
    \end{overpic}
    \phantomsubcaption\label{fig:linewidth_vs_pressure}
  \end{subfigure}
  \hfill
  \begin{subfigure}[b]{0.45\textwidth}
    \centering
    \begin{overpic}[width=\textwidth, height=0.8\linewidth]{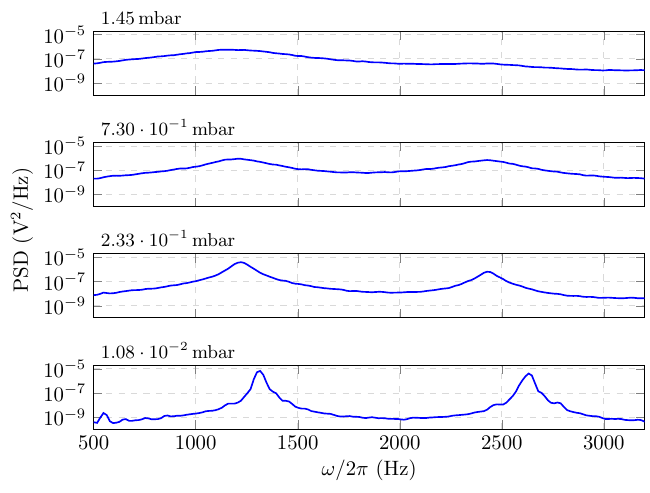}
      \put(98,74){\makebox(0,0)[tr]{\textbf{(b)}}}
    \end{overpic}
    \phantomsubcaption\label{fig:psd_vs_pressure}
  \end{subfigure}
    \caption[Mass determination via pressure scan]{Mass determination via pressure scan: (a) The motional linewidth ($\gamma_{\text{gas}}$) as a function of pressure. The linear fit allows extraction of the particle's radius. (b) A \gls{psd} at different pressures. Two of the motional modes become increasingly sharp as pressure is reduced. The second ($x$) radial mode is visible only at lower pressures due to its proximity in frequency to the first ($y$) radial mode.}
    \label{fig:mass_measurement}
\end{figure}

The particle mass ($m$) was extracted by measuring the motional linewidth, $\gamma_{\text{gas}}$, as a function of background gas pressure (see Fig.~\ref{fig:linewidth_vs_pressure} for an example) \cite{bullier_characterisation_2020}. Using this method, a preliminary measurement for a levitated \gls{nd} indicated a radius of \qty{91}{\nano\meter}, which corresponds to a mass of \qty{9.6e-18}{\kilogram}. Fig.~\ref{fig:psd_vs_pressure} illustrates how the motional modes become more resolved as the pressure decreases.

\subsection{Long-Term Frequency Stability}
\label{sec:frequency_stability}

\begin{figure}[H]
    \centering
    \includegraphics[width=0.8\textwidth]{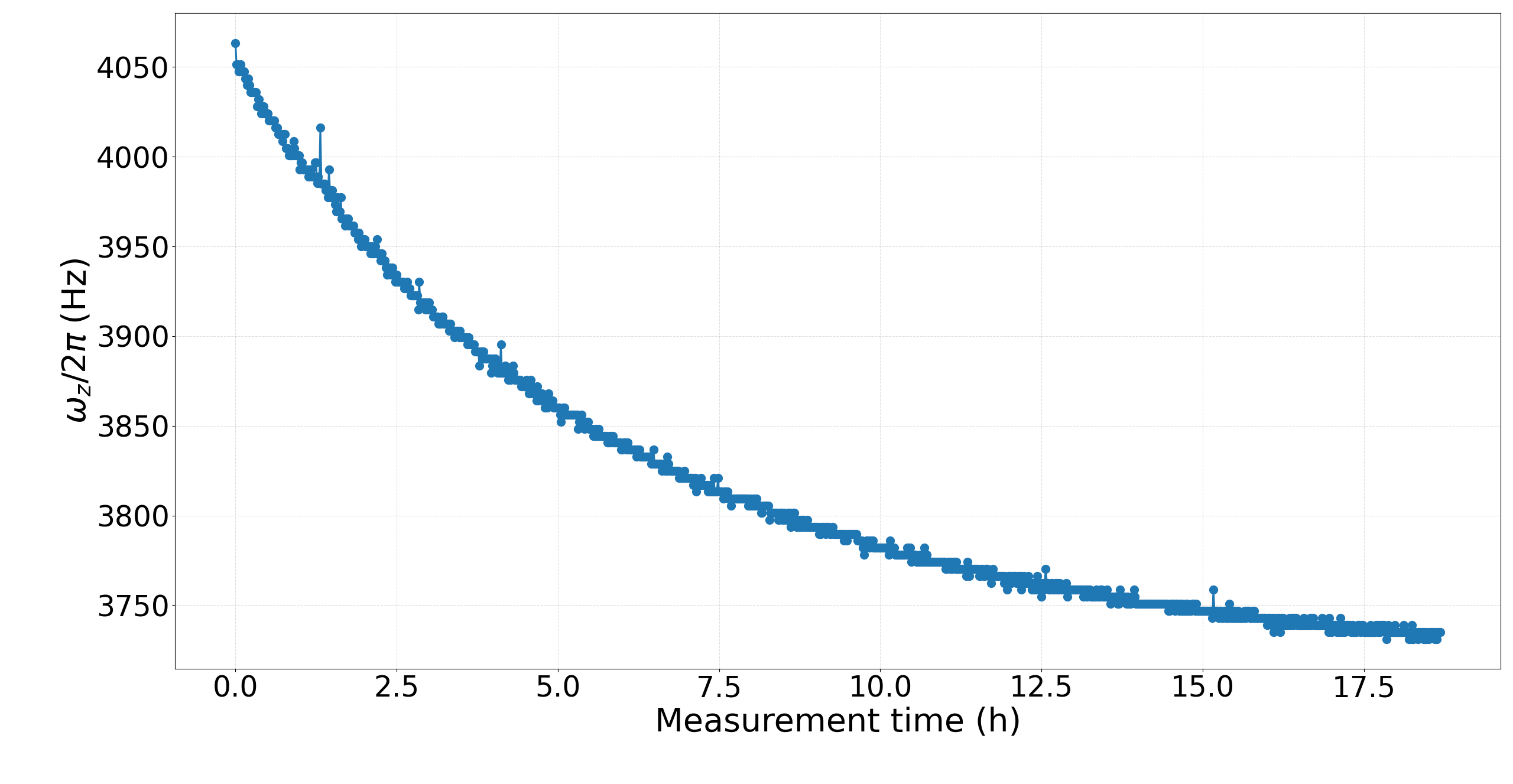}
    \caption[Axial frequency time trace]{A time trace of the axial secular frequency, showing a decay over a period of 18 hours. This decay is attributed to charging effects on nearby dielectric surfaces as explained in the main text.}
    \label{fig:freq_drift}
\end{figure}

The long-term stability of the trapping potential was monitored by tracking the secular frequencies over time. We observe a slow, continuous drift in the axial secular frequency over a period of 18 hours immediately following the \gls{esi} loading process (Fig.~\ref{fig:freq_drift}). This instability is attributed to stray electric fields generated by charge accumulating on nearby dielectric surfaces during the electrospray process \cite{bullier_characterisation_2020}, which slowly alter the trapping potential. Since our \gls{esi} is inside the vacuum chamber, even a short operation sprays many charges that stick to the dielectric surfaces of the trap and slowly dissipate. We see evidence that longer operations of the \gls{esi} correlate with stronger decay in the secular frequency until the frequency stabilizes.

\subsection{Active Feedback Cooling}
\label{sec:cooling_results}

We implemented and demonstrated active feedback cooling of one axis of radial motion (y-axis) of a levitated \gls{nd} at a pressure of \qty{8.0e-5}{\milli\bar}. For the cooling, the detection optical setup was modified, so that all the forward scattering signal along with the transmitted beam went to only one BPD -- bypassing the DM and both beam-splitters -- to increase the \gls{snr}. The D-shaped mirror was oriented horizontally to provide better detection of the particle's motion along y-axis. An electrical feedback signal, phase-locked to the particle's motional signal from a balanced detector, was applied to the cooling electrode at the bottom of the trap to damp the motion along y-axis. The feedback cooling loop is implemented via an IQ filter integrated in \gls{rp} and accessed using the PyRPL \cite{Neuhaus2024PyRPL} package. The IQ filter is used as an adjustable narrow bandpass filter and as a \gls{pll} to lock onto the phase of the motional mode and generate a feedback force (detailed procedures are described in Appendix \ref{appendix:cooling}). By optimizing the phase and gain of the feedback signal, we achieved a reduction in the motional energy of the particle.

\begin{figure}[H]
  \centering
  \begin{subfigure}[b]{0.51\textwidth}
    \centering
    \begin{overpic}[width=1.15\linewidth,keepaspectratio]{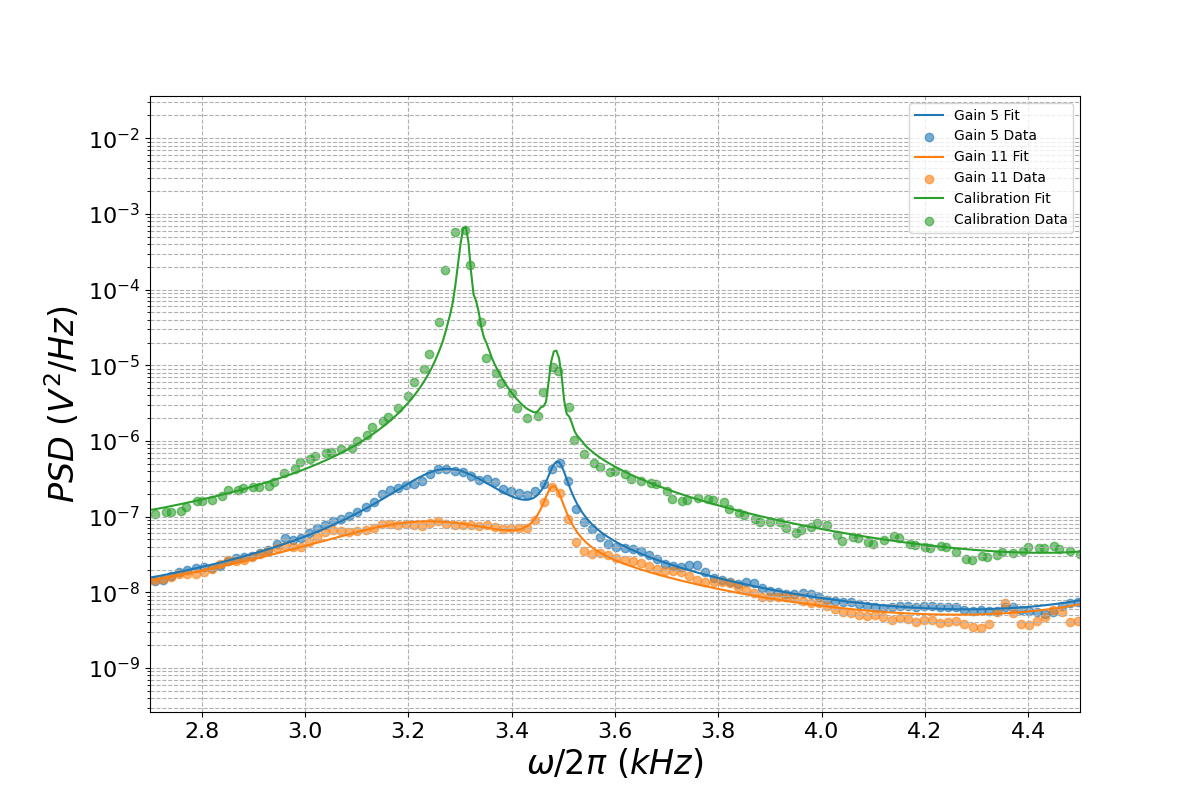}
      \put(20,57){\makebox(0,0)[tr]{\textbf{(a)}}}
    \end{overpic}
    \phantomsubcaption\label{fig:cooling_psd}
  \end{subfigure}
  \hfill
  \begin{subfigure}[b]{0.44\textwidth}
    \centering
    \begin{overpic}[width=\linewidth, height=0.8\linewidth]{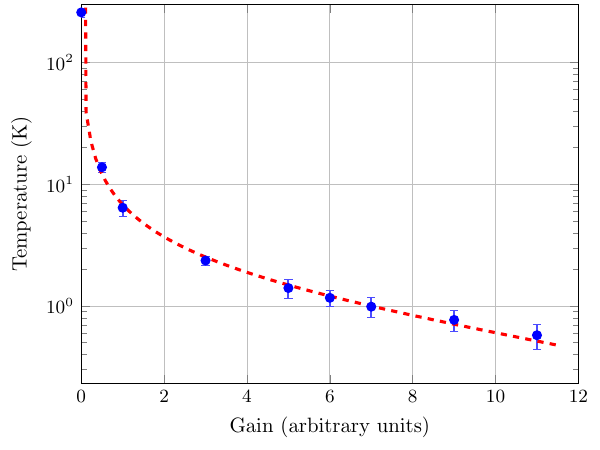}
      \put(24,77){\makebox(0,0)[tr]{\textbf{(b)}}}
    \end{overpic}
    \phantomsubcaption\label{fig:temperatures_vs_gain}
  \end{subfigure}
    \caption[First cooling results]{First cooling results: (a) PSD of the cooling at gains 5 and 11 and the calibration trace. Cooling of the two radial modes obtained at \qty{8.0e-5}{\milli\bar}, calibration trace obtained at \qty{1.62e-2}{\milli\bar}. The feedback signal operates on the lower (y) frequency of motion but effectively cools both modes. (b) Temperature of y-radial motion achieved at different cooling gains. The red dotted line is a guide-to-the-eye fit. The error bars represent the difference between two methods of calculating the area of the PSD, one by numerical integration of the raw data, and the second by using the area under the fitted curve.}
    \label{fig:cooling}
\end{figure}

Fig.~\ref{fig:cooling_psd} shows examples of \gls{psd}s at different feedback gains, along with the calibration signal that serves as the reference for thermometry of the motional temperature as described in \cite{Hebestreit2018CalibrationEnergy}. The calibration trace was recorded right before the main cooling sequence, at pressure \qty{1.62e-2}{\milli\bar}, where the particle is in thermal equilibrium with the room-temperature gas (free-molecular/Epstein regime).

Fig.~\ref{fig:temperatures_vs_gain} shows result temperatures of y-radial motion for different feedback gains. The lowest temperature achieved is $570\pm100\,\qty{}{\milli\kelvin}$, which to the best of our knowledge is comparable to the state-of-the-art: \qty{1}{\kelvin} for a ND in a Paul trap \cite{jin_quantum_2024}, and \qty{600}{\milli\kelvin} for a microdiamond cluster in a magneto-gravitational trap \cite{Hsu2016CoolingMagnetoGravitational}.

The cooling is limited by the low \gls{na} of the forward detection ($\mathrm{NA} = 0.25$). Moving to backward detection, where the large \gls{na} lens ($\mathrm{NA} = 0.67$) can collect more of the scattered light, would enable better cooling. Other detection schemes can also be implemented to generate better \gls{snr} and enable lower temperatures \cite{dania_position_2022, werneck_cross_correlation_2024}.

\subsection{UHV Trapping}
We trapped a \gls{nd} in a \qty{1e-8}{\milli\bar} vacuum environment, which, to our knowledge, is the deepest vacuum in which a \gls{nd} has been trapped in a Paul trap \cite{jin_quantum_2024}. \Glspl{nd} have been trapped down to \qty{1e-9}{\milli\bar}, but in a magneto-gravitational trap and with the optics outside the chamber \cite{DUrso_private_2025}. According to \cite{Pino2018OnChipSuperconductingMicrosphere}, the decoherence rate of an interferometer due to scattering of gas molecules is
\begin{equation}
    \gamma_a=\frac{16 \pi \sqrt{2 \pi}}{\sqrt{3}} \frac{P R^2}{\sqrt{3 m_a K_b T_e}},
    \label{eq:Decoherence_gas}
\end{equation}
where $T_e$ is the environmental temperature, $K_b$ is the Boltzmann constant, $m_a$ is the mass of a gas molecule, and $P$ is the environmental pressure. As we strive toward the first \gls{sgi} time of \qty{100}{\micro\second} with a \qty{40}{\nano\meter} diameter \gls{nd}, the minimum pressure to allow this is $P=\qty{6e-8}{\milli\bar}$. Trapping a \gls{nd} in a deeper vacuum is another step to achieve before the implementation of the full \gls{sgi}.

\subsection{High Intensity Illumination}

\begin{figure}[H]
    \centering
    \includegraphics[width=0.8\linewidth]{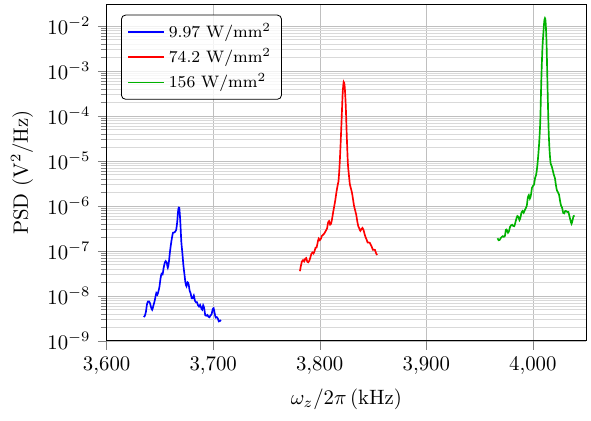}
    \caption[Intensity results]{The \gls{psd} of the axial motion of a levitated \gls{nd} for different laser intensities. As we increase the intensity, the noise floor rises and the trapping frequency rises.}
    \label{fig:intensity_psd}
\end{figure}

As previously reported \cite{rahman_burning_2016}, we expect the \gls{nd} to burn or undergo graphitization and be lost from the trap when illuminated with a relatively high-intensity laser. We started the pressure reduction while illuminating the \gls{nd} with low intensity (\qty{0.637}{\watt\per\square\milli\meter}) in order to avoid losing the trapped \gls{nd}. When we reached a vacuum of \qty{1e-8}{\milli\bar}, we then increased the \qty{1560}{\nano\meter} intensity and expected to observe a particle loss event. However, we increased the intensity of the trapping laser up to \qty{165}{\watt\per\square\milli\meter} and observed no evidence of graphitization, with the particle remaining trapped. We observe an increase in both the \gls{snr} and the trap frequency as we increase the intensity of the detection laser, as can be seen in Fig.\,\ref{fig:intensity_psd}. The source of the latter correlation, if indeed it is a correlation, is still not clear to us.

\section{Discussion}
\subsection{Technical Achievements and Significance}
Our demonstration of stable \gls{nd} trapping at \qty{1e-8}{\milli\bar} represents an advancement towards the implementation of \gls{sgi}. This \gls{uhv} environment is crucial for future matter-wave interferometry applications, as it substantially reduces decoherence from gas molecule collisions according to Eq.~\ref{eq:Decoherence_gas}. The achieved pressure is well below the theoretical threshold of \qty{6e-8}{\milli\bar} required for our target short-duration interferometer of \qty{100}{\micro\second}.

\subsection{High-Intensity Laser detection}

\begin{figure}[H]
    \centering
    \includegraphics[width=0.8\textwidth]{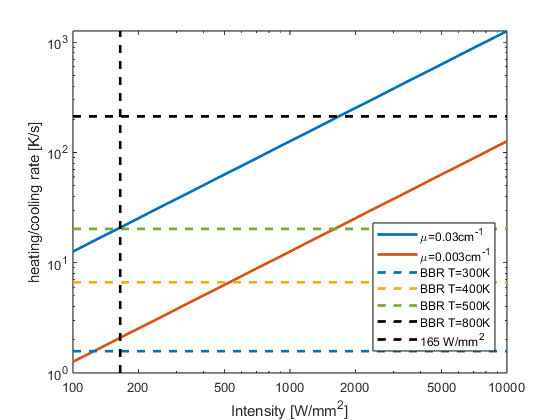}
    \caption[]{Heating from laser radiation vs. cooling from \gls{bbr} of a levitated \gls{nd}. Two absorption coefficients of \qty{0.03}{\per\cm} and \qty{0.003}{\per\cm} for standard and low absorption grade \gls{nd} and the equation dominating the dynamics are taken from \cite{frangeskou_pure_2018}. For standard absorption, at the \qty{165}{\watt\per\square\milli\meter} intensity introduced, the two processes balance at \qty{500}{\kelvin}.}
    \label{fig:heating_vs_blackbody}
\end{figure}

Even at ”short” interferometry times of \qty{100}{\micro\second}, the \gls{nd} \gls{sgi} must operate at \gls{uhv} conditions of \qty{1e-8}{\milli\bar}. This was assumed to be a critical obstacle, since it is well known that in dipole traps the NDs graphitize and are lost from the trap below a few mbar \cite{rahman_burning_2016}. In earlier levitation experiments in Paul traps, the NDs were also lost from the trap, or the NV contrast was reduced after CW illumination at \qty{532}{\nano\meter} \cite{Delord2017DiamondsPaulTrapNVHeating}. However, Jin et al. \cite{jin_quantum_2024} measured the internal temperature of a levitated \gls{nd}, which remains stable at approximately 350\,K at pressures below \qty{1e-5}{\milli\bar}. This result suggests that operating a \gls{sgi} under UHV conditions is realistic. Our demonstration of stable ND trapping at \qty{1e-8}{\milli\bar} confirms this and represents an advancement towards the implementation of a \gls{sgi}.

We analyze the heating/cooling rates and estimate that detection at \qty{1560}{\nano\meter} does not considerably heat the \gls{nd}, so that the properties of the \gls{nv} will remain stable (see Fig.~\ref{fig:heating_vs_blackbody}).

\subsection{Charge Dynamics and Stability}
The measured charge-to-mass ratios of \qtyrange{10}{75}{\coulomb\per\kilogram} span a wide range, with consistent ability to select a certain region by choosing the initial trapping conditions. The observed frequency drift over \qty{18}{\hour} (Fig.~\ref{fig:freq_drift}) is an expected result but presents a challenge as every new particle trapping event requires a few days relaxation of the system (sometimes up to 3 days, depending on the amount of spraying before a trapping event occurs). This drift, attributed to charge accumulation on dielectric surfaces during the electrospray process \cite{bullier_characterisation_2020}, suggests that alternative trapping methods would be preferable for future work. Options include a dry source for \gls{nd} similar to that presented in \cite{bykov_direct_2019} or like we are developing \cite{benjaminov2025uhvloading}, or a different \gls{esi} mechanism that completely isolates the sprayed solvent from the trapping area. 

\subsection{Cooling Performance and Limitations}
Our cooling results show a temperature of $570\pm100\,\qty{}{\milli\kelvin}$ in the y axis radial mode, a significant step towards the temperatures required for the \gls{nd} \gls{sgi} experiment. Our simulations show that for a short-duration \gls{nd} \gls{sgi} a \gls{com} temperature of 0.5\,mK is good enough. We expect to achieve this in the very near future, e.g., with the higher NA lens (i.e., moving from NA=0.25 to 0.67, see Fig.\,\ref{fig:Lens holder with labels}). The high-frequency trap we developed is also expected to significantly help reach low temperatures \cite{Skakunenko_Needle_trap}. We do not think this will be necessary, but if so, additional detection and cooling techniques will be considered, such as cavity-enhanced cooling \cite{delic2020cooling, Dania2025HighPurityQuantumOptomechanics}, Kalman filter assisted particle tracking \cite{magrini_real_time_2021}, or other detection and cooling approaches \cite{kamba_revealing_2023, kremer_all_electrical_2024} combining feedback and electrical cooling.

\section{Outlook}
The successful trapping of a \gls{nd} at \qty{1e-8}{\milli\bar} under high-intensity laser illumination represents a significant step toward realizing quantum superposition interferometry with levitated NDs, as for a short-duration ND SGI this vacuum level is sufficient. 

In the immediate term, our primary focus will be on achieving better cooling of the levitated \gls{nd} to reduce its \gls{com} motion to the required 0.5\,mK temperature. We will later on work to achieve even lower temperatures required for long-duration interferometry.  Concurrently, we aim to measure an \gls{odmr} spectrum from the levitated \gls{nd}, which will confirm the preservation of the \gls{nv} centers' quantum properties under our trapping conditions.

While our current vacuum level of \qty{1e-8}{\milli\bar} is sufficient for short-duration interferometry (several tens of microseconds), longer interferometer durations will require deeper vacuum conditions. Our long-term objectives, therefore, include pursuing \gls{uhv} conditions reaching pressures below \qty{1e-11}{\milli\bar}. This will require baking procedures without removing the optics and a UHV compatible source, as we are indeed developing \cite{Givon_ND_fabrication, benjaminov2025uhvloading}.

Beyond translational motion control, we aim to achieve control over all rotational degrees of freedom of the levitated \gls{nd}. We plan to develop and implement techniques for measuring and cooling the rotational motion, which is necessary for the complete control over the external degrees of freedom and represents a significant challenge in levitated optomechanics. Our theoretical work on rotation cooling \cite{Muretova_ND_theory, Japha2023RotationsSGI} shows that this is feasible. Using our high-frequency Paul trap \cite{Skakunenko_Needle_trap} is expected to assist with CoM and rotation cooling.

The results shown in this paper were obtained using a collection lens that has a relatively small \gls{na} compared to the focusing lens. Future experiments will be conducted using backscattered detection, which features significantly higher \gls{na} optics, thereby increasing the amount of scattered light collected and providing better control over the ratio between scattered and unscattered laser light. This will, in turn, improve the precision of particle position measurements and the \gls{snr}. Additionally, a third collection lens with an even higher \gls{na} positioned at a different angle will enable efficient \gls{nv} center spectroscopy and coherent manipulation of the quantum states within the levitated \gls{nd}. See our work on NV spectroscopy in \cite{Levi_Quantum_control_NV}.
Last, before the ND SGI sequence begins we will neutralize the particle to avoid strong decoherence (see our work on neutralization in 
\cite{Liran_ND_neutralization}).
We emphasize that while all elements such as the Paul trap electrodes and current-carrying wires for the diamagnetic trap and SG forces will eventually be integrated onto a chip, we initially plan to use our existing setup for the first interferometry experiments, where a chip with the current-carrying wires will be brought with a piezo stage to the vicinity of the ND.
\section*{Acknowledgments}
We thank Markus Aspelmeyer and his team for useful discussions. Maria Muretova thanks the Ariane de Rothschild Women Doctoral Program for outstanding female PhD students for its support. We thank the BGU Atom-Chip Group support team, especially Menachem Givon, Zina Binstock, Dmitrii Kapusta and Yaniv Bar-Haim for their support in building and maintaining the experiment. Funding: This work was funded by the Gordon and Betty Moore Foundation (doi.org/10.37807/GBMF11936),
and Simons Foundation (MP-TMPS-00005477).

\appendix
\appendixpage

\section{Decoherence from Blackbody radiation} \label{appendix:Gravitationlly_incduce_quantum_stateBR}

\begin{figure}[H]
    \centering
    \includegraphics[width=0.8\textwidth]{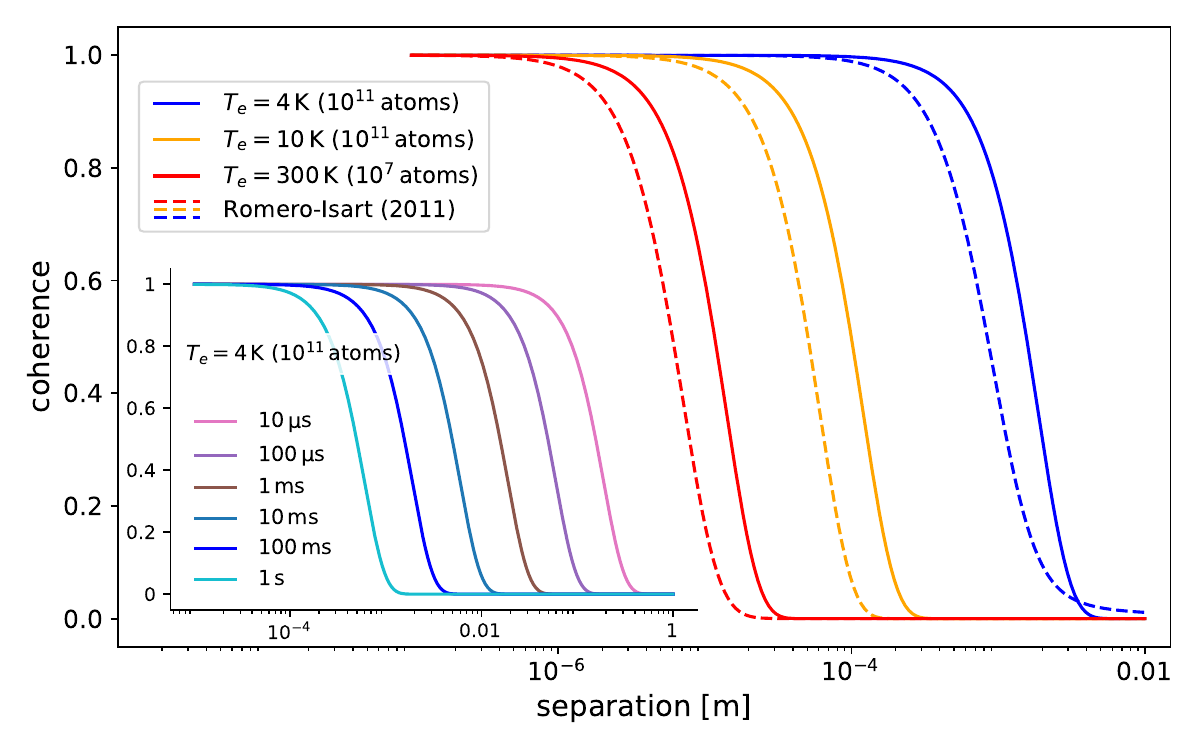}
    \caption[]{Coherence drop due to \gls{bbr} as a function of maximal wavepacket separation \cite{RomeroIsart2011CollapseModels, Bateman2014NearFieldInterferometry, Chang2010CavityOptomechanicsLevitatedNanosphere}. The main plot shows calculations for two models: dashed curves from \cite{RomeroIsart2011CollapseModels}, solid curves from \cite{HenkelFolman2024UniversalLimitPhonons}. The predictions for 300\,K are for a 100$\,\mu s$ duration experiment with $10^7$ atoms, and it can be seen that no cold environment (chip or chamber) is needed up to a 10$\,\mu m$ splitting. The curves for 10\,K and 4\,K are computed for a 0.1\,s duration experiment with $10^{11}$ atoms, showing that a splitting of up to 1\,mm is possible for cryogenic temperatures. The inset illustrates that at long times, the accumulated effect of decoherence is nonzero (model of \cite{HenkelFolman2024UniversalLimitPhonons}), although the splitting is below the coherence length of \gls{bbr} radiation ($\sim 1\,{\rm mm}$ at 4\,K). This can be attributed to many scattered photons. The calculations were done by Carsten Henkel.}
    \label{fig:blackbody radiation}
\end{figure}

Environmental decoherence is induced by any coupling of the environment to the quantum system. This may be seen as noise imparted by the environment onto the quantum system, increasing the uncertainty in the delicate quantum phases, or as entanglement between the quantum system’s states and the states of the environment giving rise to orthogonality. An omni-present type of coupling to the environment is due to \gls{bbr}. \gls{bbr} photons can be absorbed by the ND or scattered by it. Similarly, the environment can absorb \gls{bbr} photons emitted by the ND. There are several models describing this process, which do not completely agree with each other. It would be very interesting to learn more about this process through the ND SGI. In Fig.~\ref{fig:blackbody radiation} we plot the theoretical predictions of ND SGI as a function of separation and temperature. We do so first for the simpler experiment with a light $10^7$-atoms ND at room temperature, and for the heavier $10^{11}$-atoms ND at cryogenic temperature. We also show the effect of time, where one can see that at \qty{100}{\milli\second} the limit of the spatial splitting is about the same as the mean wavelength of the \gls{bbr}. This could be understood to be the “resolution” with which the environment is probing our spatial superposition.

\section{Testing gravitationally-induced quantum state
reduction}  \label{appendix:Gravitationlly_incduce_quantum_state}

By pushing the limits of the above roadmap for generating and monitoring superpositions of nanodiamonds consisting of many billions of atoms, we aim to enable novel tests of fundamental physics. In this appendix, we discuss the possibility of using this scenario to check for gravitationally-induced quantum state reduction (GQSR), as proposed by Diósi \cite{Diosi1987UniversalMasterEquation, Diosi1989UniversalReduction}, Penrose \cite{Penrose1986GravitySVR, Penrose1996GravityRoleQSR, Penrose1998QuantumComputationEntanglement, Penrose2014GravitizationQM}, and others (see \cite{Oppenheim2023PostquantumClassicalGravity, Howl2019BECUnification, Oppenheim2023GravInducedDecoherenceVsDiffusion} and references therein). According to these proposals, one does not view the superposition principle of quantum mechanics as valid at scales clashing with, e.g., the equivalence principle of general relativity (this approach is termed "gravitizing quantum theory", in contrast to the more standard approach of "quantizing gravity"). As a result, a superposition of a particle larger than the Planck mass ( $\sim 2 \times 10^{-8} \mathrm{~kg}$ ) may be considered impossible \cite{Feynman1957ChapelHillComments}. Due to the involvement of time and distance scales as well, the proposals can be checked with much smaller masses.

Specifically, the idea is that the lifetime of a superposition of two possible positions for a massive particle is of the order of $\hbar / E_{G}$, where $E_{G}$ is the gravitational self-energy of the mass distribution corresponding to the difference between the mass distributions of the two branches of\\
the superposition. For a nano-object of spherical shape and uniform density in a superposition of two positions separated by a distance $d$, Penrose finds that \cite{Penrose2014GravitizationQM}

\begin{equation*}
E_{G}=\frac{G M^{2}}{R} f\left(\frac{d}{2 R}\right) \tag{12}
\end{equation*}

where $R$ is the radius of the sphere of mass $M$, and

\[
f(\lambda)= \begin{cases}2 \lambda^{2}-1.5 \lambda^{3}+0.2 \lambda^{5} & 0 \leq \lambda \leq 1  \tag{13}\\ 1.2-0.5 \lambda^{-1} & \lambda \geq 1\end{cases}
\]

Note that when the separation $d$ is much less than the diameter of the sphere, $f$ is small and quadratic in $d$; when it is equal to the diameter we have $f=0.7$ and it does not grow dramatically for larger separations, approaching a maximum of $f=1.2$ as the separation increases to infinity (alternative shapes of the superposed particle were also considered, with largely similar results \cite{Howl2019BECUnification}).

A reasonable choice of parameters for the QG nanodiamond setup would be a mass of $10^{-15} \mathrm{~kg}$, corresponding to $10^{11}$ carbon atoms (see Table I). For a separation of $d=2  \,\mathrm{\mu m}$ between the branches of the superposition, one obtains $f \simeq 1$ and a lifetime of order $\hbar R / G M^{2} \simeq 0.6 \mathrm{~s}$. Thus, if the different sources of decoherence applicable to such a massive nanodiamond can be controlled to the degree that the associated lifetimes would be on the order of seconds, GQSR should be observable. Since the mass appears quadratically in this expression (and $R$ grows with $M$ in a very modest manner), increasing it can substantially shorten the timescale for GQSR.

A disadvantage of this setup is that the prediction of the lifetime for GQSR is not sharp. If interference were still observed for times larger than $\hbar R / G M^{2}$ by some finite factor, one could still tune a parameter in the Di\'{o}si-Penrose proposals to accommodate the result and postpone GQSR to later times \cite{Ghirardi1990CSLGravity}. Conversely, if decay of the superposition is observed on just the proposed timescale, one could be concerned that some conventional source of decoherence is operative.

The calculations in this part were done by Nathan Argaman.

\section{Cooling procedure and code}\label{appendix:cooling}
To identify the effect every feedback electrode has on each of the motional modes, we couple a "tickler" RF voltage at the resonance frequency and observe the detected signal of the particle as it heats up and the peak becomes driven by the tickler. Then we can easily estimate the voltage needed in order to cool or heat the particle on each electrode for each axis of motion. While it depends on the distance of the electrode from the particle, resonant excitation at the secular frequency produces observable increases in PSD peak amplitude (on the order of \qty{10}{\deci\bel}) with voltages that are typically $10$ to $100$ times smaller than those required to achieve similar spectral changes through off-resonant driving.
After calibrating for the needed feedback voltage to start cooling, we start the cooling process. The feedback loop is implemented, as mentioned in the main text, using a \gls{rp} \gls{fpga} board with the use of the PyRPL package. The package contains easily programmable access to multiple hardware modules inside the \gls{rp}, specifically 3 IQ modules implemented in each of the \glspl{rp} can be used in parallel to cool each of the motional modes.
An IQ module (In-phase/Quadrature module) is a digital signal processing component that performs lock-in amplification by demodulating a signal into its in-phase (I) and quadrature (Q) components relative to a reference frequency. This allows for precise extraction of amplitude and phase information from noisy signals, making it ideal for feedback cooling applications where detecting small motional signals is crucial.
The configurable output ports allow multiple modes to be cooled simultaneously on a single electrode, or alternatively, enable parametric feedback cooling of rotational modes—currently the most effective approach for rotational cooling \cite{gao_feedback_2024}. The code for using the \gls{rp} IQ filter allows control of the gain, the bandwidth of the filter, and the relative phase between the detection signal and the filtered feedback signal. We tune the filter frequency to the secular oscillation mode, carefully scan the phase to find the best cooling phase then adjust the gain to slowly reduce the effective temperature of the particle. A short and specific part of the code showing the use of a single IQ to cool a single frequency is added here.

\definecolor{vscodegreen}{RGB}{106, 153, 85}
\definecolor{vscodeblue}{RGB}{86, 156, 214}
\definecolor{vscodeorange}{RGB}{206, 145, 120}
\definecolor{vscodepurple}{RGB}{197, 134, 192}
\definecolor{vscodeyellow}{RGB}{220, 220, 170}
\begin{lstlisting}[
language=Python,
backgroundcolor=\color{black!5},
basicstyle=\ttfamily\footnotesize,
commentstyle=\color{vscodegreen},
keywordstyle=\color{vscodeblue}\bfseries,
stringstyle=\color{vscodeorange},
numberstyle=\tiny\color{gray},
identifierstyle=\color{black},
emphstyle=\color{vscodeblue},
emph={frequency,bandwidth,gain,phase,acbandwidth,amplitude,input,output_direct,quadrature_factor},
showstringspaces=false,
breaklines=true,
frame=single,
rulecolor=\color{gray!30},
numbers=left,
stepnumber=1,
numbersep=8pt,
tabsize=4,
captionpos=b,
caption={IQ module configuration for feedback cooling using PyRPL}
]
--- Connection & Configuration ---
#Set the IP address of your Red Pitaya board.
RED_PITAYA_IP = "0.0.0.0"  # Replace with your actual Red Pitaya IP address.
#Define a name for your Pyrpl configuration file.
CONFIG_NAME = "Axial_mode"
#--- Hardware & Signal Routing ---
#Choose the physical input channel ('in1' or 'in2') connected to the detector.
INPUT_CHANNEL = 'in1'
#--- Lock-in Amplifier (IQ Module) Settings ---
#The expected resonant frequency of the levitated particle in Hertz (Hz).
FREQUENCY = 6.168e3
#The band-pass filter bandwidth in Hz. 
FILTER_BANDWIDTH_HZ = 200
#The gain applied to the demodulated signal before output.
OUTPUT_GAIN = 12
#The phase shift in degrees applied to the reference oscillator. To measure velocity,
DEMODULATION_PHASE_DEG = 270
#The bandwidth of the AC coupling filter in Hz. This is used to remove any DC offset
#from the sensor signal, focusing only on the particle's oscillations.
AC_COUPLING_BANDWIDTH_HZ = 150
#The amplitude of the IQ module's internal signal generator. For feedback cooling,
#this is set to 0.0, as we are only use the module as a feedback generator. and do not want
#to inject any additional signal.
INTERNAL_OSCILLATOR_AMPLITUDE = 0.0
#Determines which signal is sent to the module's output_signal multiplexer to be used with #other.
#modules. 'output_direct' uses the raw demodulated signal, while 'quadrature' would give #the low
#pass filtered demodulated signal.
OUTPUT_DIRECT_ROUTING = 'output_direct'
#The output signal routing for the I/Q demodulation module. This can be 'off', 'out1', #out2', or 'both'
DIRECT_OUTPUT_ROUTING = 'out1'
#The quadrature factor for the I/Q demodulation module. This is not used for feedback #cooling,
#so it is set to 0.
QUADRATURE_FACTOR = 0
=============================================================================
Main Script Logic
=============================================================================
#Import the main Pyrpl class to control the Red Pitaya.
from pyrpl import Pyrpl
#Create a Pyrpl object to connect to the Red Pitaya using the defined parameters.
p = Pyrpl(config=CONFIG_NAME, gui=False, hostname=RED_PITAYA_IP)
#Access the RedPitaya object for low-level hardware control.
r = p.rp
#Select the specified I/Q demodulation module from the list of available modules.
iq = r.iq0 
#Configure the I/Q demodulation module using the general parameters defined above.
#This setup creates a real-time filter to isolate the particle's velocity signal.
iq.setup(frequency=FREQUENCY,
bandwidth=FILTER_BANDWIDTH_HZ,
gain=OUTPUT_GAIN,
phase=DEMODULATION_PHASE_DEG,
acbandwidth=AC_COUPLING_BANDWIDTH_HZ,
amplitude=INTERNAL_OSCILLATOR_AMPLITUDE,
input=INPUT_CHANNEL,
output_direct=DIRECT_OUTPUT_ROUTING,
quadrature_factor=QUADRATURE_FACTOR)
\end{lstlisting}


\end{document}